\documentclass[a4paper,11pt]{article}
\usepackage{pos}
\usepackage{siunitx}
\usepackage[capitalise]{cleveref}
\usepackage{wrapfig}
\usepackage{enumitem}

\title{AugerPrime Surface Detector Electronics: requirements, verification and performance}
 \ShortTitle{AugerPrime electronics}

\author*[a]{Martina Boh\'a\v cov\'a}

\affiliation[a]{Institute of Physics, Prague, Czech Republic}

\onbehalf{for the Pierre Auger Collaboration$^b$}
\affiliation[b]{Observatorio Pierre Auger, Av.\ San Mart{\'\i}n Norte 304, 5613 Malarg\"ue, Argentina\\
Full author list: {\rm\url{https://www.auger.org/archive/authors_icrc_2025.html}}}

\emailAdd{spokespersons@auger.org}

\abstract{The Pierre Auger Observatory has recently undergone a major upgrade, called AugerPrime, tailored to answer the current most pressing questions in the ultra-high-energy cosmic ray (UHECR) detection. The AugerPrime upgrade consists of the addition, on top of each station, of a scintillator detector, to separate the muonic and electromagnetic component of the shower for better primary identification, and of a radio detector to measure the emission of air showers in 30-80 MHz range. An additional small diameter photomultiplier is installed in each station to increase the dynamic range for signal detection. New electronic modules, installed on all stations, provide a sufficient number of channels for the readout of the additional detectors, as well as faster sampling, increased dynamic range and processing capability. In this contribution we summarize the performance of the new electronic modules with respect to the requirements, describe the verification procedure and give the results in the laboratory tests compared to the performance in the field.}

\ConferenceLogo{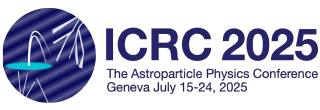}
\FullConference{%
39th International Cosmic Ray Conference (ICRC2025)\\
15 -- 24 July, 2025\\
Geneva, Switzerland}

\begin{document}
\maketitle
\section{The Piere Auger Observatory and the AugerPrime upgrade}

The Pierre Auger Observatory features a~large detection area
collecting unprecedented event statistics, and a~combination of several
detection techniques that allows one to lower the systematic uncertainties of
the measurements. The Surface
detector (SD) array is formed by 1600 water-Cherenkov detectors (WCD)
placed in a~triangular grid with \SI{1500}{\metre} spacing (SD-1500) on the area
of around \SI{3000}{\km\squared}. The SD-1500 array is fully efficient above
\SI{3e18}{\electronvolt} and continuously samples the extensive air showers (EAS) generated by the interaction of primary UHECRs with atmospheric nuclei,
with a duty cycle of nearly \SI{100}{\percent}. A nested region within the SD-1500 array contains additional 61 WCDs in half distances (SD-750) and even at 433 m distances (SD-433), covering an area of about 27 km$^2$ and 2 km$^2$ respectively. This allows us to lower the energy threshold of the Observatory below 10$^{17}$ eV. The WCDs are overseen by 27
fluorescence telescopes located along the boundary of the array,
which observe the longitudinal profile of
the air showers. The operation of the fluorescence detector (FD) is limited to 
clear moonless nights. A~thorough description of the Observatory is
given in~\cite{auger_nim}.

After nearly 20 years of operation, the Observatory has been
upgraded. The main scientific aspects motivating the upgrade are the
following: understand the origin of flux suppression, identify the sources of UHECRs,  solve the discrepancy between hadronic interaction models and the measured shower parameters, and look for possible effects beyond the standard model. The upgrade aims at resolving these
issues by obtaining more information on the mass composition of cosmic rays at the highest energies. 
Each WCD has been equipped with a~\SI{4}{\metre\squared}
plastic scintillator mounted on the top (Surface Scintillator Detector
or SSD). The two detectors will provide complementary information
about the electromagnetic and muonic components of the shower. An
additional Radio Detector (RD) is mounted on top of each WCD to observe the radio signals in the 30~--~\SI{80}{\MHz} band from
inclined showers to add yet another measurement of the electromagnetic
component. A~small PMT (SPMT), which is accompanying the three large PMTs from the baseline design inside the WCD, further extends the dynamic range and allows one to
study signals closer to the shower core. Finally, underground muon detectors (UMD) are deployed near 61 stations of the denser area (SD-750 and SD-433) to provide a direct measurement of the muonic component of the EAS. A~more powerful, modernized electronics was installed in each WCD interfacing the additional detectors together with the original ones. A detailed description of the SD electronics upgrade can be found in \cite{sdeu_jinst} and is summarized below. A~comprehensive description of the Observatory upgrade
can be found in \cite{PDR}.

The deployment of the pre-production and production electronics, together with SPMTs and SSDs, started in mid-2020 and was completed in July 2023. Commissioning studies are in progress since December 2020, to monitor various performance parameters. The installation of the RD detectors started in mid-2023 and was completed by the end 2024. The installation of UMD is well advanced and is expected to be completed in 2025.

\section{Requirements and design of the Surface Detector electronics upgrade}
\begin{figure}
\centering
\def\h{0.34}
\includegraphics[height=\h\textwidth]{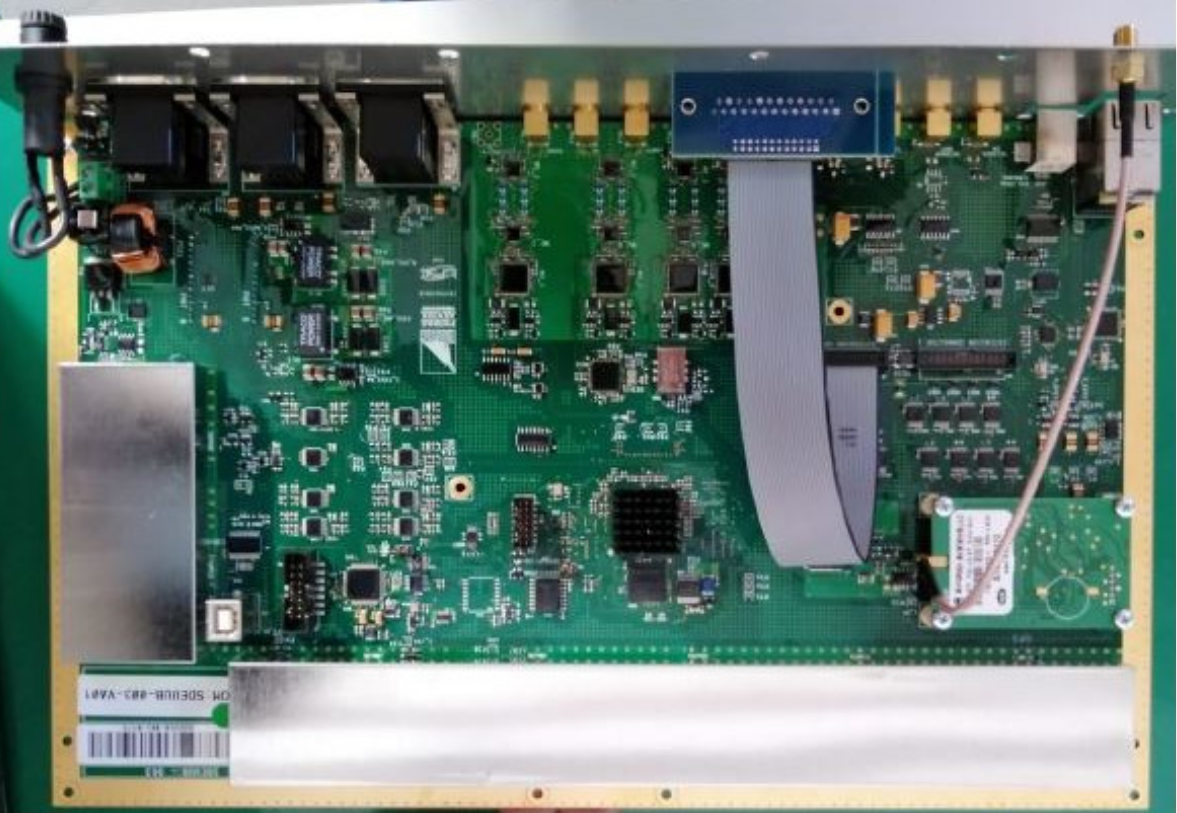}\hfill
\includegraphics[height=\h\textwidth]{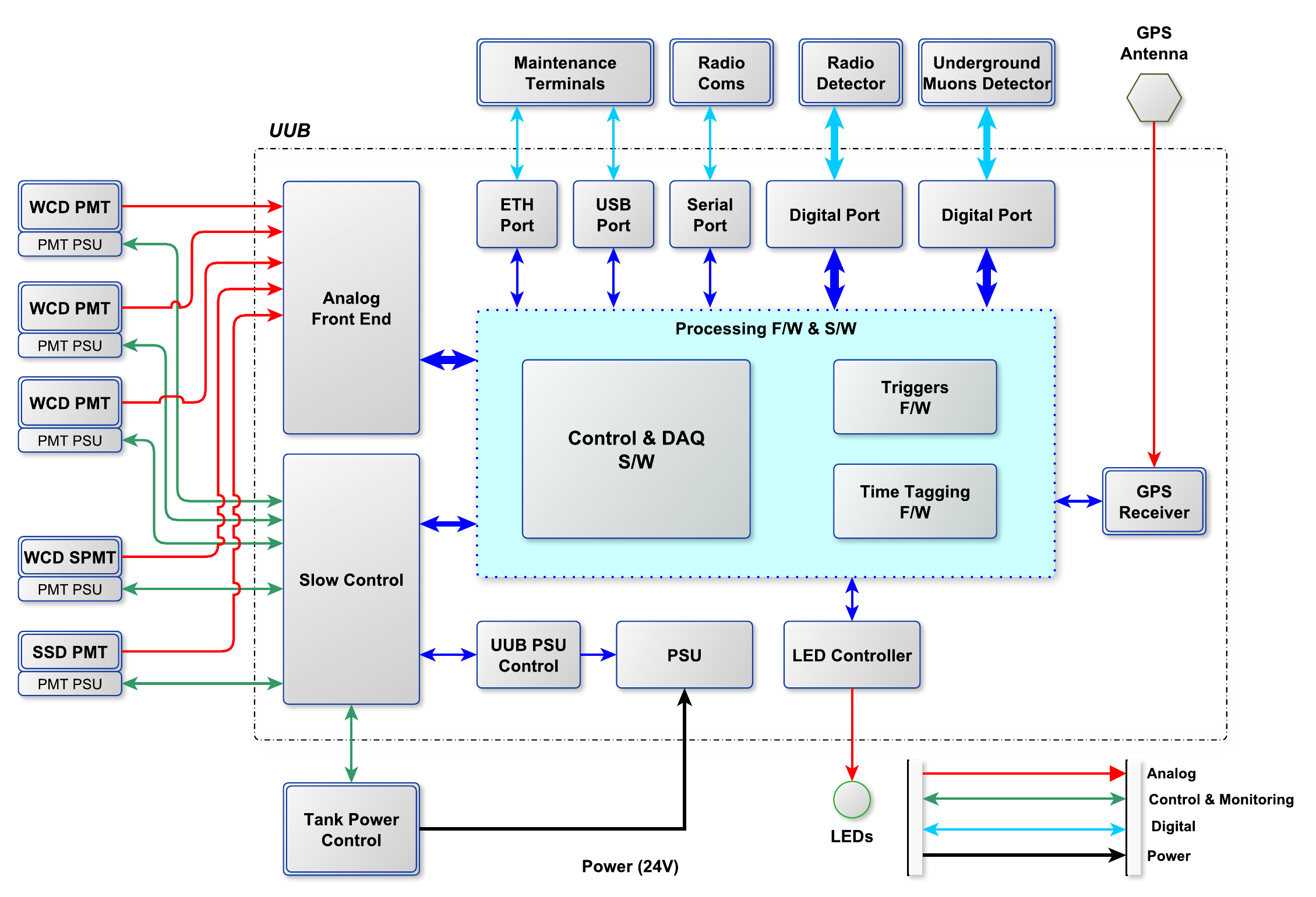}
\caption{\emph{Left:} UUB including front panel, GPS receiver and shielding covers. 
\emph{Right:} Electrical interfaces of the UUB.}
\label{fig:UUB}
\end{figure}

The general requirement for the electronics upgrade was to increase the data quality exploiting faster
sampling for Analog-Digital-Converter (ADC) traces, better timing accuracy, increased dynamic
range, enhanced local trigger and processing capabilities, more powerful local station processor with a Field Programming Gate Array (FPGA), and improved calibration and monitoring capabilities. Backwards-compatibility was assured by retaining the time span
of the PMT traces and providing for digital filtering and down-sampling of the traces to emulate the previous triggers in addition to new ones, implemented in the FPGA firmware. The most important functional and configurational design features are: digitization of the anode signals from the PMTs at a sampling frequency of 120 MS/s with a resolution of 12 bit and 10 ADC analog input channels (for the three large PMTs and the SSD PMT with two gains each plus the SPMT and a spare channel).
Event time tagging with a resolution of 5 ns and a stability better than 5\% depending on temperature variations is achieved by implementing Synergy SSR-6TF timing GPS receivers. 
All the functions (except GPS receivers) are integrated on a single Upgraded Unified Board (UUB) including also power-supply unit with safety features. The total power consumption has to be kept on a minimum level and it currently amounts to about 14 W including all the interfaces. UUB is designed to fit into the preexisting RF-enclosure and cable connections are compatible with the previous electronics board (UB). The UUB architecture is designed with a Xilinx Zynq FPGA containing two embedded ARM Cortex A9 333 MHz microprocessors. The FPGA is connected to a 4 Gbit LP-DDR2 memory and a 1 Gbit Flash memory. The FPGA implements all basic digital functions such as the read-out of the ADCs, the generation of triggers, and the integration of GPS receiver, clock tree and memories.
Two digital ports on the UUB enable communication with additional detector systems.
The requirement of total RMS integrated noise at the ADC output below 0.5 LSB (Least Significant Bit) for the LG channel and 2 LSB for the HG channel is achieved by a careful analog design and the use of low noise dual ADC (AD9628). 
A powerful 16-bit RISC CPU ultra-low-power  micro-controller (MSP430) is used for the PMT high-voltage control, the supervision of various supply voltages and monitoring of environmental sensors.

\subsection{Production, testing, verification and deployment}

\begin{figure}
\centering
\def\h{0.30}
\includegraphics[height=\h\textwidth]{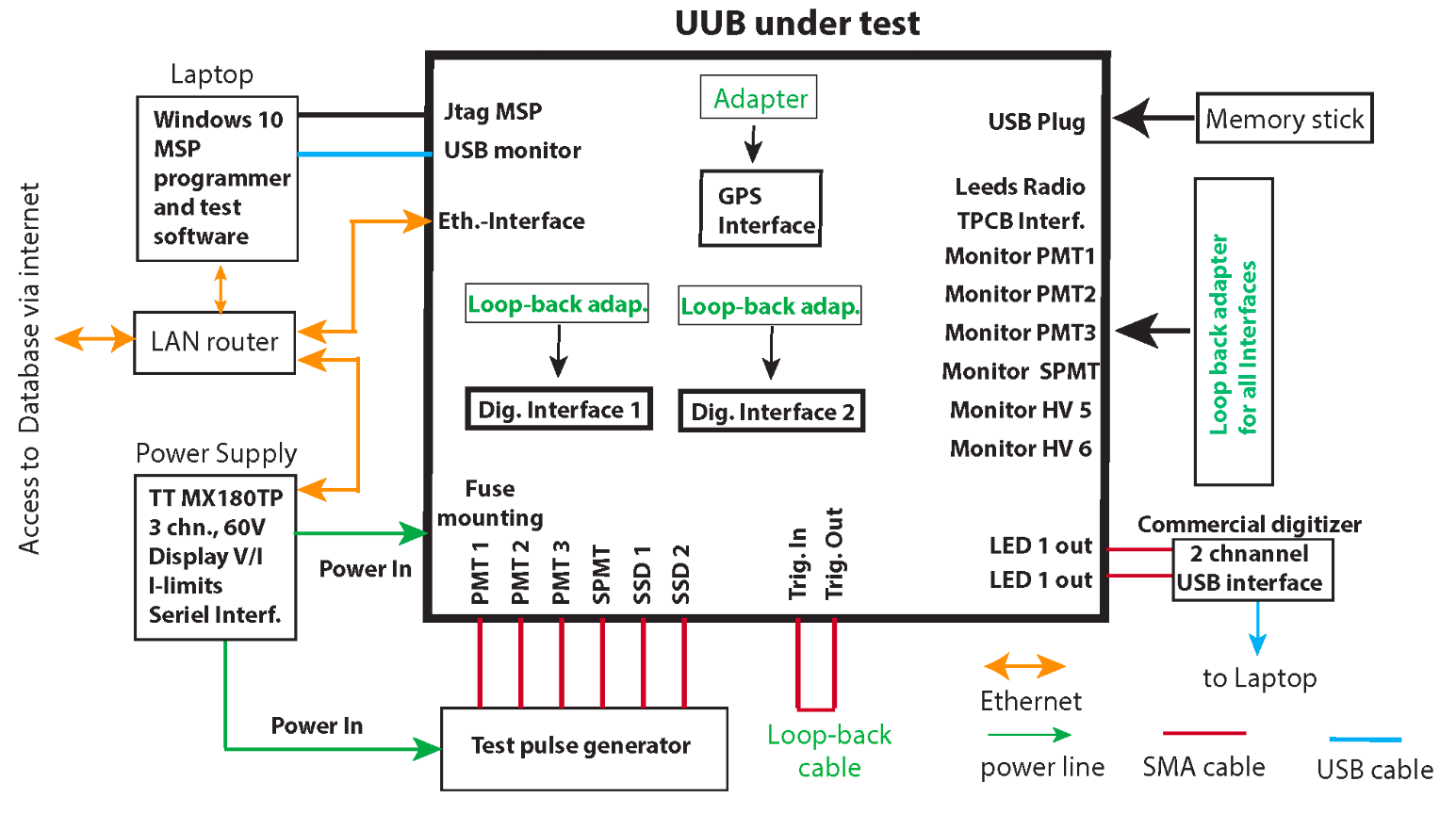}\hfill
\includegraphics[height=\h\textwidth]{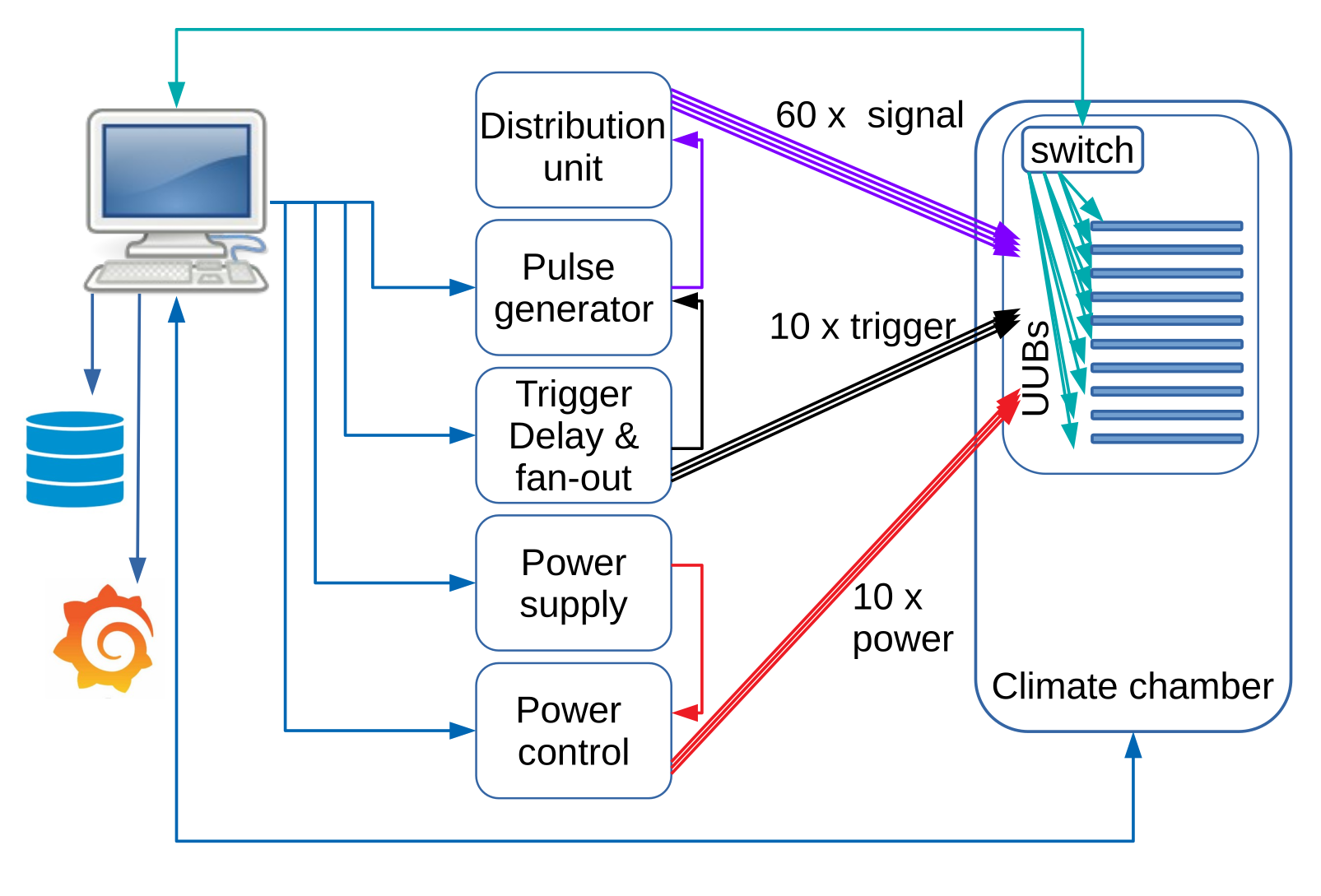}
\caption{\emph{Left:} Manufacturing test bench scheme. 
\emph{Right:} ESS test bench scheme.}
\label{fig:test_bench}
\end{figure}

The UUBs were manufactured by the A4F company (Angel4Future, Bari, Italy). Quality control and testing of the boards was performed at a test benches provided by the Auger Collaboration.
As a first step of the testing procedure, the UUBs were subjected to an automatic optical inspection to detect any missing components or soldering problems. The subsequent manufacturing test verifies all electrical lines and the basic functionality of the UUB.  During the manufacturing test, an updated firmware for the MSP microcontroller and FPGA was installed . The scheme of the manufacturing test bench is shown in the left panel of \cref{fig:test_bench}.

The UUBs were then submitted to an Environmental Stress Screening (ESS), to characterize the behavior of the new electronics under changing environmental conditions typically observed at the Observatory site (see \cref{fig:gain} the right panel for the typical temperature profile inside the electronics enclosure of WCD) and to provoke early failures. During the automated procedure a set of ten  UUBs at a time were first subjected to burn-in (rapid temperature changes for 24 hours) with noise, baselines and temperature regularly monitored. In a second step, 10 temperature cycles, from -20$^\circ$C to +70$^\circ$C (temperature change of 3$^\circ$C/min) were applied. The performance of each UUB was monitored at five different temperatures, in particular studying the
noise, baseline and linearity dependence on temperature, the stability of the ADCs and the anti-alias filter, and a verifying of the over/under voltage protection.
The scheme of the ESS test bench is shown in the right panel of \cref{fig:test_bench}, and a
detailed description of the test procedure can be found in \cite{MartinaICRC19}.

Several failures were encountered during the test procedure, such as ADC flipping bits, ADCs or baseline instabilities,  or failures of the 3.3V converter, mainly due to nonconforming components and production faults. Most of these failures were mitigated by repairing and component exchange, yielding a final failure rate at production of about 1.3\%.

After ESS, the UUBs were transported to the Pierre Auger Observatory, where they underwent a final verification and a full functionality test before deployment in the field.  First, a visual inspection was performed to search for transportation damages. The UUBs were then assembled adding the GPS receiver, cables, connectors, front panel, and the complete setup was placed inside a protective RF-enclosure. The final end-to-end verification was performed before the whole assembly was taken to the field and integrated into each surface detector station. 

\section{Calibration}
\begin{figure}
\centering
\def\h{0.35}
\includegraphics[height=\h\textwidth]{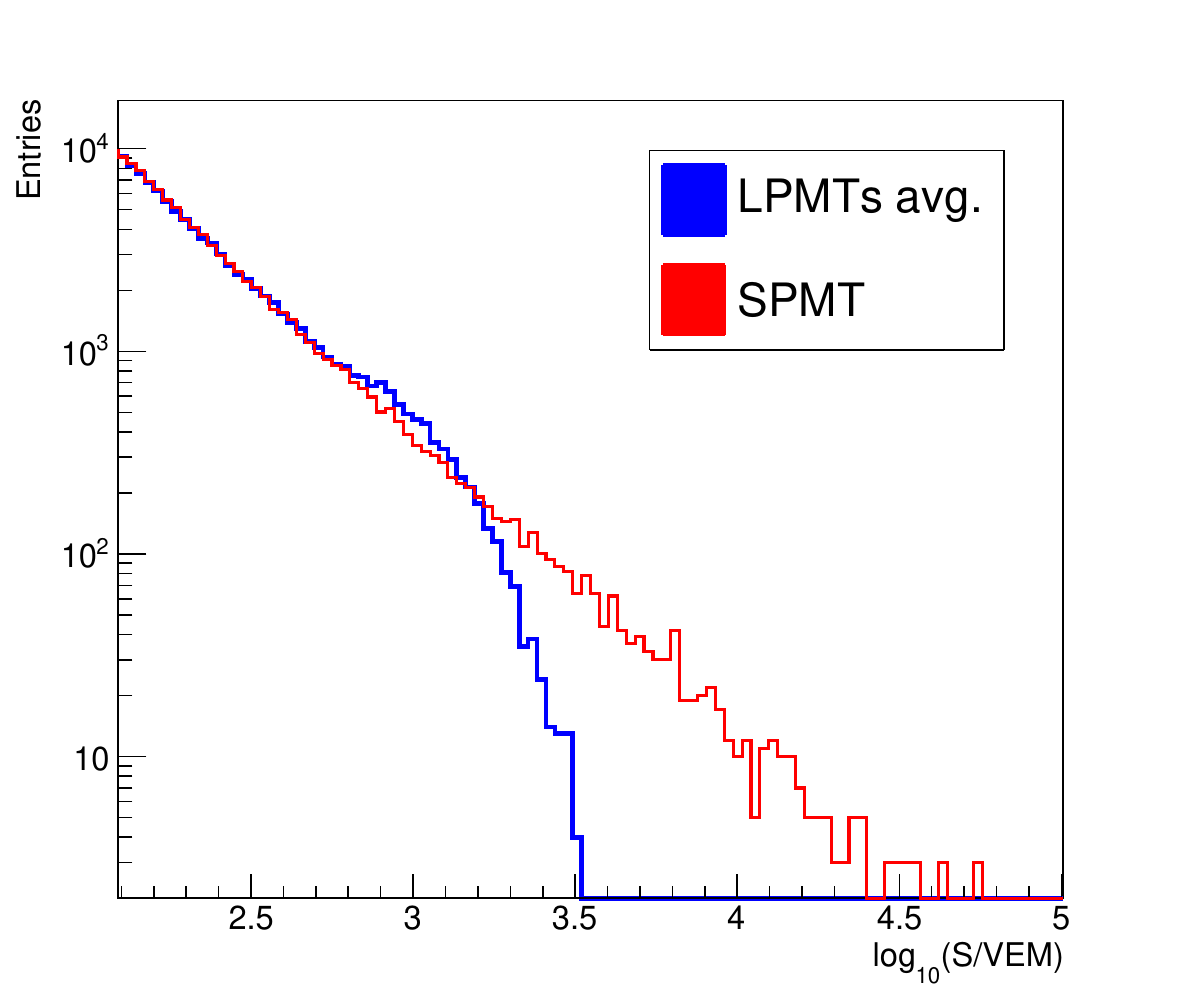}
\includegraphics[height=\h\textwidth]{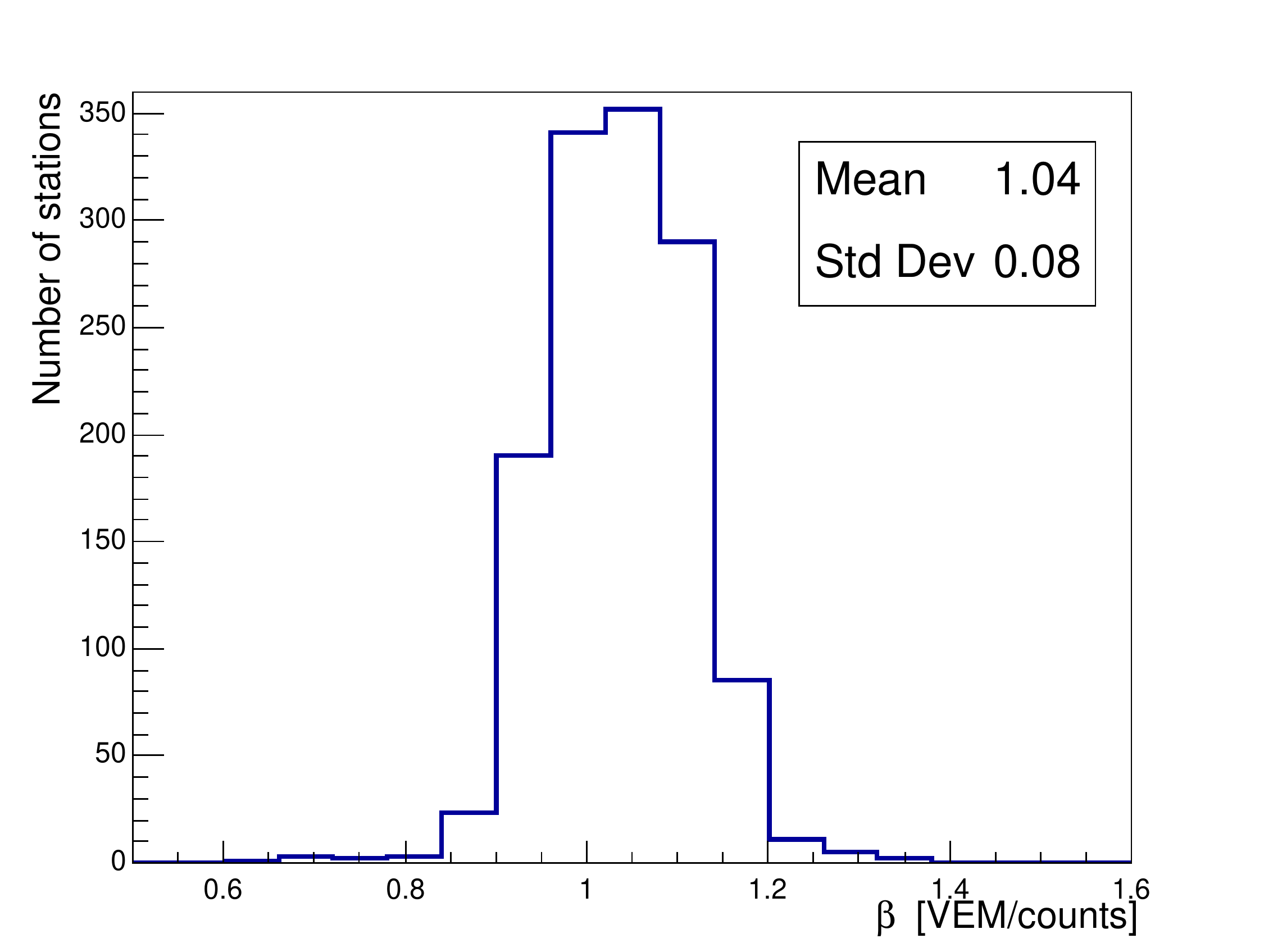}
\caption{\emph{Left:} Extension of the WCD dynamic range to 20 000 VEM using the small PMT. 
\emph{Right:} Distribution of the average cross-calibration factors $\beta$ on a single day, after the installation and setting of all the SPMTs in the array. }
\label{fig:SPMT}
\end{figure}

\subsection{WCD and SSD calibration}
Both detectors are continuously calibrated with background atmospheric muons. The charge histograms produced each minute show characteristic humps corresponding to the Vertical Equivalent Muon (VEM) in the WCD and the Minimum Ionizing Particle (MIP) in the SSD. The mean measured charge values are about 1400 and 110 ADC channels, respectively for the VEM and MIP. In addition,  40\% of the muons are measured by WCD and SSD in coincidence (corresponding to the overlapping area), which helps to significantly reduce the background and makes the VEM and MIP humps clearly distinguishable. Since this calibration is done every minute of the data taking, any long term temperature dependence of various components is taken into account during analysis of the shower data.

\subsection{Cross-calibration of the SPMT}
Limited by their photocathode dimension, the new small PMTs cannot be calibrated exploiting the same method as for the large PMTs, as only about one photoelectron per muon is recorded by the SPMT with respect to about 90 collected in the LPMT. For this reason, a new method was developed based on the minimization of the differences between the calibrated signal spectrum of the large WCD PMTs and that of the small PMT. The so-called $\beta$ factor derived in this way is then used to cross-calibrate the SPMT, converting the collected charge measured in ADC channels into a signal in VEM. A dedicated selection of signals from local low energy showers is exploited, measured separately in each station at a rate of about 200\,events/hour. The minimization is performed in a superposition region, where the signal is large enough for the SPMT, but not saturated in LPMTs. The logarithm of the charge spectrum in one upgraded WCD station is shown in the left panel of \cref{fig:SPMT}. Thanks to the SPMT, the dynamic range is extended to more than 20 000 VEM. In order to follow the daily evolution of the SPMT gain due to the temperature variations, the cross-calibration is performed in 8-hour sliding windows. In this way, a precision in the determination of the cross-calibration factor $\beta$ of ${\sim}2.2\%$ is obtained.
 The stability of the $\beta$ factor across the whole array is shown in the right panel of \cref{fig:SPMT}. A yearly evolution ot about $\pm5$\% is expected, due to the seasonal variation of the average temperature in the field. Further details on the SPMT implementation and performance can be found in \cite{GialexICRC23}.

\begin{figure}
\centering
\def\h{0.34}
\includegraphics[height=\h\textwidth]{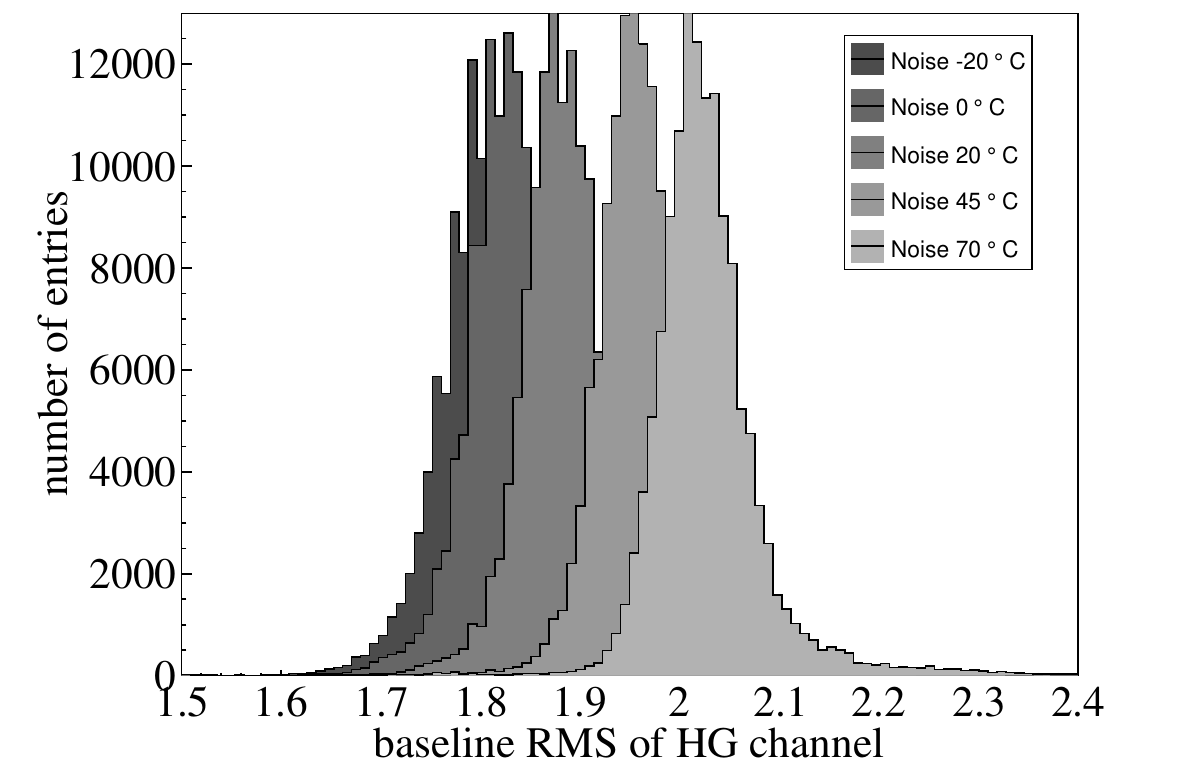}\hfill
\includegraphics[height=\h\textwidth]{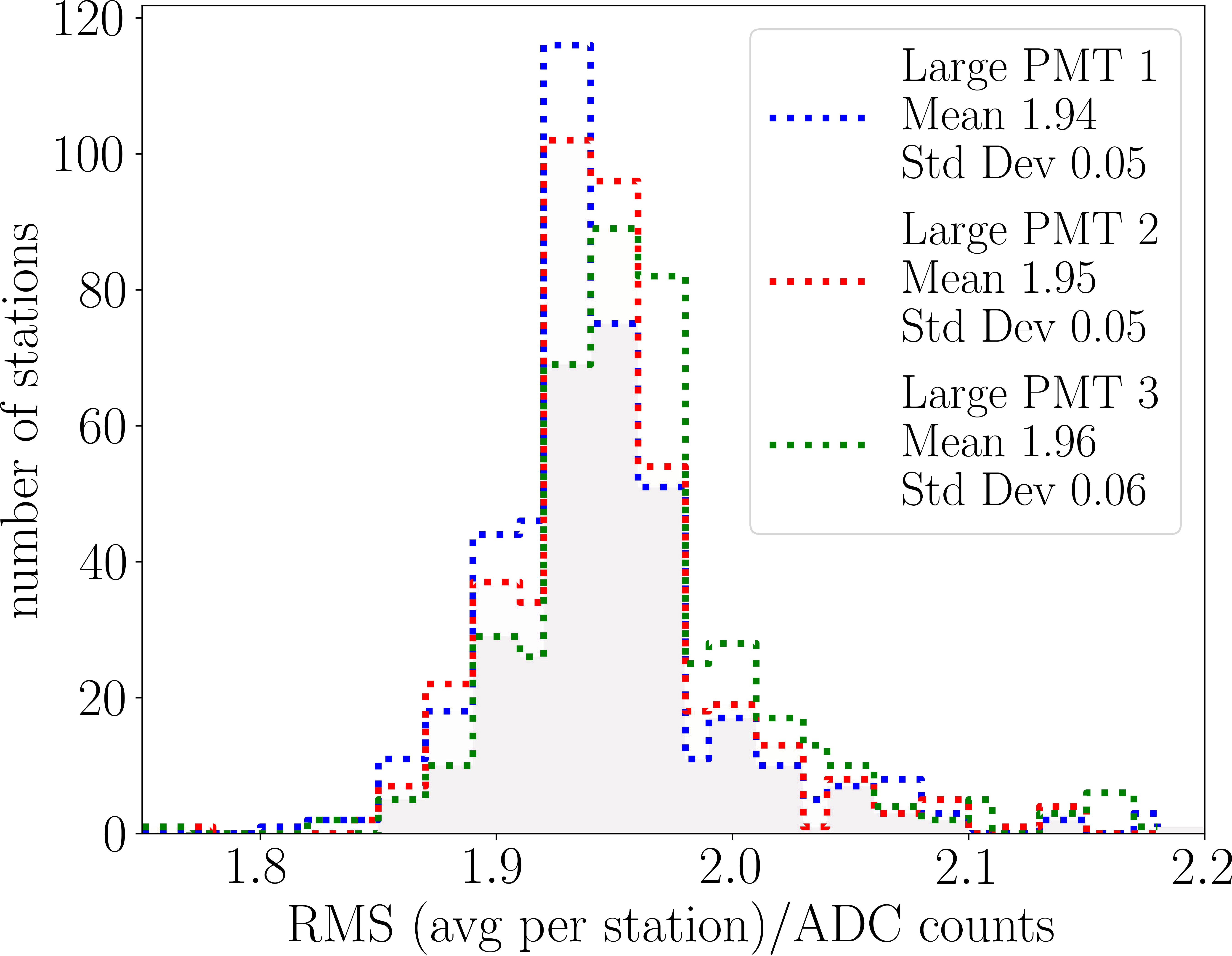}
\caption{\emph{Left:} Noise measured on the UUB  high gain channels at the ESS test bench. All three Large PMTs for all UUBs are plotted together.
\emph{Right:} Noise of the HG channels of the individual PMTs as measured in the field.}
\label{fig:noise}
\end{figure}

\section{Performance}
\subsection{Noise levels}
The noise performance of each channel was verified in the 
ESS test bench. For the high-gain channel the noise was measured to slightly increase with temperature and to be below 2 ADC bins for most channels (as seen in the left panel of \cref{fig:noise}), meeting the requirements. The SSD high-gain channels show even better noise performance for all temperatures. The noise of the SPMT and the low-gain channels of
LPMTs and SSD PMTs were measured to be around 0.5 ADC channel, again in agreement with
the requirements. During field operation, the electronic modules are exposed to external environmental factors, affecting the connections to PMTs, the power system, and communication and timing systems, which can
induce noise. Also the grounding quality, depending on field conditions, can impact the noise performance of the electronics. Compared to the previous UB electronics, the UUB has a factor 4 lower thresholds and is more susceptible to external noise. Sudden bursts of noise were observed in the traces on daily bases close to sunrise and were found to be connected with the battery charging. Modification of the tank power control board mitigated this problem. Enhanced battery maintenance, together with better cabling organization and better grounding, further improved noise levels in the field. 
Large spikes in trigger rates were observed
in coincidence with distant lightning, which resulted in a data acquisition crash. This effect was tracked down to bi-polar signals observed in some traces. The problem was eliminated by a trigger conditioning algorithm applied in the FPGA before triggering, which removes bi-polar oscillations in the baseline while preserving
the unipolar physics signal.

\subsection{Time resolution}
The upgraded electronics uses SSR-6TF timing GPS receivers to synchronize the detectors. These receivers are functionally compatible with the previous electronics and their intrinsic accuracy given by the manufacturer is 2 ns. The timing performance of the SSR-6FF GPS receiver was verified in the laboratory, relative to an FS275 GPS-disciplined rubidium atomic clock. The absolute
timing accuracy was found to range from 2.3 ns (over timescales of a few seconds) up to
about 6 ns (over timescales of several hours). The relative timing accuracy between two receivers was measured in a climate chamber across a temperature range relevant for the electronics enclosure in the field, and reached 2.1 ns. The verification of the timing performance of the GPS SSR-6TF receivers mounted on UUBs after deployment in the field
was done using a synchronization cable to send timing signals between
two closely positioned ($\sim 20$ m) WCDs, yielding a timing accuracy of about 5 ns, consistent with the lab measurements and the timing granularity as implemented on the UUB.
Similar measurement performed using real showers reached an accuracy of 13 ns, which is, however, dominated by shower-to-shower fluctuations.


\subsection{Trigger performance}
A detailed description of the trigger implementation can be found in \cite{sdeu_jinst}. The Observatory is currently running with compatibility triggers, that is the traces are digitally filtered using a FIR Nyquist filter with a 20 MHz cut-off, and down sampled to 40 MHz (the rate of the UB electronics), by taking every third bin, and then the trigger algorithm of the former electronics is applied. This allows us to combine data sets from the previous phase of operation and from AugerPrime during the deployment period, without disturbance to the data taking. 
 The new triggers, which take advantage of the full bandwidth, i. e. 120 MHz sampling rate of the traces are currently under development. In particular, a trigger for the Radio Detector has been developed and is currently being tested in a dedicated area of the array. The trigger performances of the previous and new electronics are compatible: the second-level trigger rates are about 20 Hz and the shower trigger rates are around 1 Hz for both types of electronics.
 The increased number of bins per trace and the additional channels result in an increased transfer time from 2 minutes to about 5 minutes per event. However, the available bandwidth is sufficient. 

\subsection{Dynamic range}

The SD signal varies from few photoelectrons for stations far from the shower core to hundreds of thousands near the shower core. To achieve this large dynamic range the LPMT signal is passing through two amplification chains implemented within the electronics, as high gain (HG) and low gain (LG) channels. 
The highest particle densities are measured by the SPMT, with an anode-channel input at a single unitary gain. A detailed account of the SPMT implementation can be found in \cite{sdeu_jinst}.

The average HG/LG ratio of the LPMT is given by the electronics design and was measured in the laboratory to be 31.7 $\pm 0.3$ with less than 1\% temperature dependence, which is fully compatible with the value obtained from suitable showers in the field, i.e. 31.7 $\pm0.4$, and represents a significant improvement in precision over the UB case, where the PMT last dynode signal was amplified and related to anode signal yielding 5\% uncertainty. The SSD anode signal is also split in two, one amplified by 32, second divided by 4 yielding a nominal ratio of 128, and covering the same dynamic range as the associated WCD. The corresponding HG/LG ratio derived from the laboratory measurements was 125.8 $\pm 1.5$, The slight discrepancy is caused by combined finite accuracies of the components.

\begin{figure}
\centering
\def\h{0.29}
\includegraphics[height=\h\textwidth]{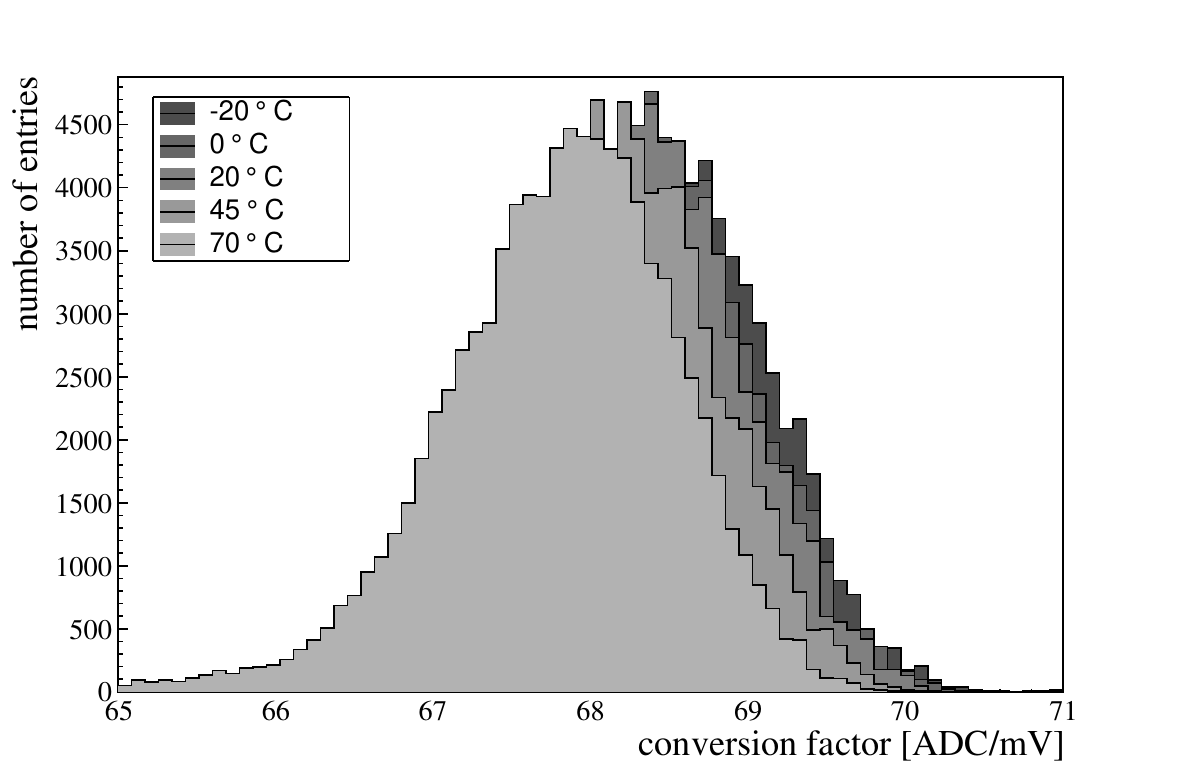}\hfill
\includegraphics[height=\h\textwidth]{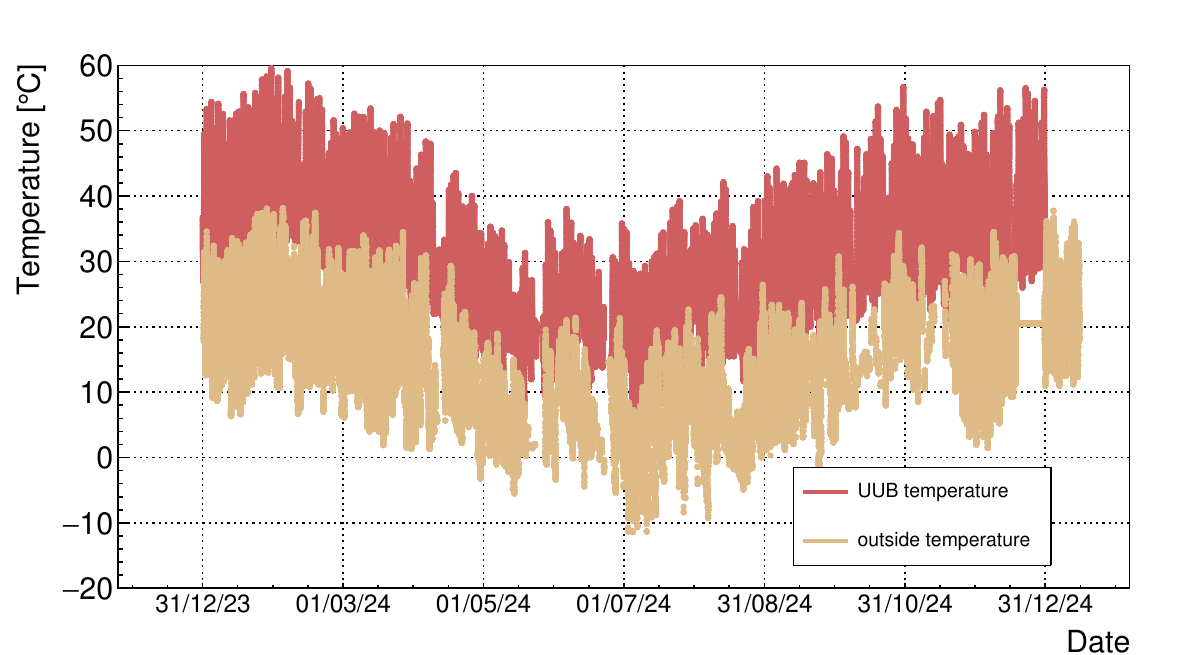}
\caption{\emph{Left:} Distribution of conversion factors from mV to ADC counts measured on the UUB high gain channels at the ESS test bench. Large PMT as well as SSD channels are included, no PMTs attached. The spread among the channels of all UUBs is larger than the temperature dependence. 
\emph{Right:} Yearly temperature evolution inside the electronics enclosure of the SD station in relation to the outside temperature. }
\label{fig:gain}
\end{figure}

\subsection{Temperature evolution}

The stability of various parameters of the front-end electronics was studied in the temperature range relevant  for the conditions in the Argentinian pampa. The temperature dependence of the gain of HG channels is plotted in \cref{fig:gain}. It can be noticed that the variation of the gain with temperature is smaller than the spread among individual UUBs and channels, when measured without PMTs. In the field, the seasonal variations of the detector response are dominated by the temperature dependence of the PMTs. It was demonstrated in the laboratory that the noise level is increasing with increasing temperature as expected. However, it stays within the requirements for most of the UUBs also at the highest tested temperature, not frequently reached in the field. 

\section*{Conclusion}

The Surface Detector electronics was upgraded to accommodate additional detectors and to enhance the experimental performances. The new features include a more powerful Xilinx Zynq-7020 FPGA for data prossing, faster ADCs sampling at 120MHz and up-to-date GPS receivers providing 5 ns timing resolution. The dynamic range of the Observatory is further increased by adding a SPMT inside the WCDs.
Extensive tests of the electronics performed in the laboratory and in the field during the commissioning phase confirm that the design meets the requirements. The two years of the AugerPrime operation show good uniformity and stable long-term performance.

\newpage

\par\noindent
\textbf{The Pierre Auger Collaboration}\\

\begin{wrapfigure}[8]{l}{0.12\linewidth}
\vspace{-2.9ex}
\includegraphics[width=0.98\linewidth]{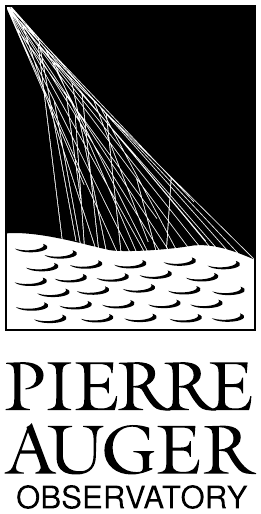}
\end{wrapfigure}
\begin{sloppypar}\noindent
A.~Abdul Halim$^{13}$,
P.~Abreu$^{70}$,
M.~Aglietta$^{53,51}$,
I.~Allekotte$^{1}$,
K.~Almeida Cheminant$^{78,77}$,
A.~Almela$^{7,12}$,
R.~Aloisio$^{44,45}$,
J.~Alvarez-Mu\~niz$^{76}$,
A.~Ambrosone$^{44}$,
J.~Ammerman Yebra$^{76}$
% ...

\end{sloppypar}

\begin{description}[labelsep=0.2em,align=right,labelwidth=0.7em,labelindent=0em,leftmargin=2em,noitemsep,before={\renewcommand\makelabel[1]{##1 }}]
\item[$^{1}$] Centro At\'omico Bariloche and Instituto Balseiro (CNEA-UNCuyo-CONICET), San Carlos de Bariloche, Argentina
\item[$^{2}$] Departamento de F\'\i{}sica and Departamento de Ciencias de la Atm\'osfera y los Oc\'eanos, FCEyN, Universidad de Buenos Aires and CONICET, Buenos Aires, Argentina
\item[$^{3}$] IFLP, Universidad Nacional de La Plata and CONICET, La Plata, Argentina
\item[$^{4}$] Instituto de Astronom\'\i{}a y F\'\i{}sica del Espacio (IAFE, CONICET-UBA), Buenos Aires, Argentina
\item[$^{5}$] Instituto de F\'\i{}sica de Rosario (IFIR) -- CONICET/U.N.R.\ and Facultad de Ciencias Bioqu\'\i{}micas y Farmac\'euticas U.N.R., Rosario, Argentina
\item[$^{6}$] Instituto de Tecnolog\'\i{}as en Detecci\'on y Astropart\'\i{}culas (CNEA, CONICET, UNSAM), and Universidad Tecnol\'ogica Nacional -- Facultad Regional Mendoza (CONICET/CNEA), Mendoza, Argentina
\item[$^{7}$] Instituto de Tecnolog\'\i{}as en Detecci\'on y Astropart\'\i{}culas (CNEA, CONICET, UNSAM), Buenos Aires, Argentina
\item[$^{8}$] International Center of Advanced Studies and Instituto de Ciencias F\'\i{}sicas, ECyT-UNSAM and CONICET, Campus Miguelete -- San Mart\'\i{}n, Buenos Aires, Argentina
\item[$^{9}$] Laboratorio Atm\'osfera -- Departamento de Investigaciones en L\'aseres y sus Aplicaciones -- UNIDEF (CITEDEF-CONICET), Argentina
\item[$^{10}$] Observatorio Pierre Auger, Malarg\"ue, Argentina
\item[$^{11}$] Observatorio Pierre Auger and Comisi\'on Nacional de Energ\'\i{}a At\'omica, Malarg\"ue, Argentina
\item[$^{12}$] Universidad Tecnol\'ogica Nacional -- Facultad Regional Buenos Aires, Buenos Aires, Argentina
\item[$^{13}$] University of Adelaide, Adelaide, S.A., Australia
\item[$^{14}$] Universit\'e Libre de Bruxelles (ULB), Brussels, Belgium
\item[$^{15}$] Vrije Universiteit Brussels, Brussels, Belgium
\item[$^{16}$] Centro Brasileiro de Pesquisas Fisicas, Rio de Janeiro, RJ, Brazil
\item[$^{17}$] Centro Federal de Educa\c{c}\~ao Tecnol\'ogica Celso Suckow da Fonseca, Petropolis, Brazil
\item[$^{18}$] Instituto Federal de Educa\c{c}\~ao, Ci\^encia e Tecnologia do Rio de Janeiro (IFRJ), Brazil
\item[$^{19}$] Universidade de S\~ao Paulo, Escola de Engenharia de Lorena, Lorena, SP, Brazil
\item[$^{20}$] Universidade de S\~ao Paulo, Instituto de F\'\i{}sica de S\~ao Carlos, S\~ao Carlos, SP, Brazil
\item[$^{21}$] Universidade de S\~ao Paulo, Instituto de F\'\i{}sica, S\~ao Paulo, SP, Brazil
\item[$^{22}$] Universidade Estadual de Campinas (UNICAMP), IFGW, Campinas, SP, Brazil
\item[$^{23}$] Universidade Estadual de Feira de Santana, Feira de Santana, Brazil
\item[$^{24}$] Universidade Federal de Campina Grande, Centro de Ciencias e Tecnologia, Campina Grande, Brazil
\item[$^{25}$] Universidade Federal do ABC, Santo Andr\'e, SP, Brazil
\item[$^{26}$] Universidade Federal do Paran\'a, Setor Palotina, Palotina, Brazil
\item[$^{27}$] Universidade Federal do Rio de Janeiro, Instituto de F\'\i{}sica, Rio de Janeiro, RJ, Brazil
\item[$^{28}$] Universidad de Medell\'\i{}n, Medell\'\i{}n, Colombia
\item[$^{29}$] Universidad Industrial de Santander, Bucaramanga, Colombia
\item[$^{30}$] Charles University, Faculty of Mathematics and Physics, Institute of Particle and Nuclear Physics, Prague, Czech Republic
\item[$^{31}$] Institute of Physics of the Czech Academy of Sciences, Prague, Czech Republic
\item[$^{32}$] Palacky University, Olomouc, Czech Republic
\item[$^{33}$] CNRS/IN2P3, IJCLab, Universit\'e Paris-Saclay, Orsay, France
\item[$^{34}$] Laboratoire de Physique Nucl\'eaire et de Hautes Energies (LPNHE), Sorbonne Universit\'e, Universit\'e de Paris, CNRS-IN2P3, Paris, France
\item[$^{35}$] Univ.\ Grenoble Alpes, CNRS, Grenoble Institute of Engineering Univ.\ Grenoble Alpes, LPSC-IN2P3, 38000 Grenoble, France
\item[$^{36}$] Universit\'e Paris-Saclay, CNRS/IN2P3, IJCLab, Orsay, France
\item[$^{37}$] Bergische Universit\"at Wuppertal, Department of Physics, Wuppertal, Germany
\item[$^{38}$] Karlsruhe Institute of Technology (KIT), Institute for Experimental Particle Physics, Karlsruhe, Germany
\item[$^{39}$] Karlsruhe Institute of Technology (KIT), Institut f\"ur Prozessdatenverarbeitung und Elektronik, Karlsruhe, Germany
\item[$^{40}$] Karlsruhe Institute of Technology (KIT), Institute for Astroparticle Physics, Karlsruhe, Germany
\item[$^{41}$] RWTH Aachen University, III.\ Physikalisches Institut A, Aachen, Germany
\item[$^{42}$] Universit\"at Hamburg, II.\ Institut f\"ur Theoretische Physik, Hamburg, Germany
\item[$^{43}$] Universit\"at Siegen, Department Physik -- Experimentelle Teilchenphysik, Siegen, Germany
\item[$^{44}$] Gran Sasso Science Institute, L'Aquila, Italy
\item[$^{45}$] INFN Laboratori Nazionali del Gran Sasso, Assergi (L'Aquila), Italy
\item[$^{46}$] INFN, Sezione di Catania, Catania, Italy
\item[$^{47}$] INFN, Sezione di Lecce, Lecce, Italy
\item[$^{48}$] INFN, Sezione di Milano, Milano, Italy
\item[$^{49}$] INFN, Sezione di Napoli, Napoli, Italy
\item[$^{50}$] INFN, Sezione di Roma ``Tor Vergata'', Roma, Italy
\item[$^{51}$] INFN, Sezione di Torino, Torino, Italy
\item[$^{52}$] Istituto di Astrofisica Spaziale e Fisica Cosmica di Palermo (INAF), Palermo, Italy
\item[$^{53}$] Osservatorio Astrofisico di Torino (INAF), Torino, Italy
\item[$^{54}$] Politecnico di Milano, Dipartimento di Scienze e Tecnologie Aerospaziali , Milano, Italy
\item[$^{55}$] Universit\`a del Salento, Dipartimento di Matematica e Fisica ``E.\ De Giorgi'', Lecce, Italy
\item[$^{56}$] Universit\`a dell'Aquila, Dipartimento di Scienze Fisiche e Chimiche, L'Aquila, Italy
\item[$^{57}$] Universit\`a di Catania, Dipartimento di Fisica e Astronomia ``Ettore Majorana``, Catania, Italy
\item[$^{58}$] Universit\`a di Milano, Dipartimento di Fisica, Milano, Italy
\item[$^{59}$] Universit\`a di Napoli ``Federico II'', Dipartimento di Fisica ``Ettore Pancini'', Napoli, Italy
\item[$^{60}$] Universit\`a di Palermo, Dipartimento di Fisica e Chimica ''E.\ Segr\`e'', Palermo, Italy
\item[$^{61}$] Universit\`a di Roma ``Tor Vergata'', Dipartimento di Fisica, Roma, Italy
\item[$^{62}$] Universit\`a Torino, Dipartimento di Fisica, Torino, Italy
\item[$^{63}$] Benem\'erita Universidad Aut\'onoma de Puebla, Puebla, M\'exico
\item[$^{64}$] Unidad Profesional Interdisciplinaria en Ingenier\'\i{}a y Tecnolog\'\i{}as Avanzadas del Instituto Polit\'ecnico Nacional (UPIITA-IPN), M\'exico, D.F., M\'exico
\item[$^{65}$] Universidad Aut\'onoma de Chiapas, Tuxtla Guti\'errez, Chiapas, M\'exico
\item[$^{66}$] Universidad Michoacana de San Nicol\'as de Hidalgo, Morelia, Michoac\'an, M\'exico
\item[$^{67}$] Universidad Nacional Aut\'onoma de M\'exico, M\'exico, D.F., M\'exico
\item[$^{68}$] Institute of Nuclear Physics PAN, Krakow, Poland
\item[$^{69}$] University of \L{}\'od\'z, Faculty of High-Energy Astrophysics,\L{}\'od\'z, Poland
\item[$^{70}$] Laborat\'orio de Instrumenta\c{c}\~ao e F\'\i{}sica Experimental de Part\'\i{}culas -- LIP and Instituto Superior T\'ecnico -- IST, Universidade de Lisboa -- UL, Lisboa, Portugal
\item[$^{71}$] ``Horia Hulubei'' National Institute for Physics and Nuclear Engineering, Bucharest-Magurele, Romania
\item[$^{72}$] Institute of Space Science, Bucharest-Magurele, Romania
\item[$^{73}$] Center for Astrophysics and Cosmology (CAC), University of Nova Gorica, Nova Gorica, Slovenia
\item[$^{74}$] Experimental Particle Physics Department, J.\ Stefan Institute, Ljubljana, Slovenia
\item[$^{75}$] Universidad de Granada and C.A.F.P.E., Granada, Spain
\item[$^{76}$] Instituto Galego de F\'\i{}sica de Altas Enerx\'\i{}as (IGFAE), Universidade de Santiago de Compostela, Santiago de Compostela, Spain
\item[$^{77}$] IMAPP, Radboud University Nijmegen, Nijmegen, The Netherlands
\item[$^{78}$] Nationaal Instituut voor Kernfysica en Hoge Energie Fysica (NIKHEF), Science Park, Amsterdam, The Netherlands
\item[$^{79}$] Stichting Astronomisch Onderzoek in Nederland (ASTRON), Dwingeloo, The Netherlands
\item[$^{80}$] Universiteit van Amsterdam, Faculty of Science, Amsterdam, The Netherlands
\item[$^{81}$] Case Western Reserve University, Cleveland, OH, USA
\item[$^{82}$] Colorado School of Mines, Golden, CO, USA
\item[$^{83}$] Department of Physics and Astronomy, Lehman College, City University of New York, Bronx, NY, USA
\item[$^{84}$] Michigan Technological University, Houghton, MI, USA
\item[$^{85}$] New York University, New York, NY, USA
\item[$^{86}$] University of Chicago, Enrico Fermi Institute, Chicago, IL, USA
\item[$^{87}$] University of Delaware, Department of Physics and Astronomy, Bartol Research Institute, Newark, DE, USA
\item[] -----
\item[$^{a}$] Max-Planck-Institut f\"ur Radioastronomie, Bonn, Germany
\item[$^{b}$] also at Kapteyn Institute, University of Groningen, Groningen, The Netherlands
\item[$^{c}$] School of Physics and Astronomy, University of Leeds, Leeds, United Kingdom
\item[$^{d}$] Fermi National Accelerator Laboratory, Fermilab, Batavia, IL, USA
\item[$^{e}$] Pennsylvania State University, University Park, PA, USA
\item[$^{f}$] Colorado State University, Fort Collins, CO, USA
\item[$^{g}$] Louisiana State University, Baton Rouge, LA, USA
\item[$^{h}$] now at Graduate School of Science, Osaka Metropolitan University, Osaka, Japan
\item[$^{i}$] Institut universitaire de France (IUF), France
\item[$^{j}$] now at Technische Universit\"at Dortmund and Ruhr-Universit\"at Bochum, Dortmund and Bochum, Germany
\end{description}

\section*{Acknowledgments}

\begin{sloppypar}
The successful installation, commissioning, and operation of the Pierre
Auger Observatory would not have been possible without the strong
commitment and effort from the technical and administrative staff in
Malarg\"ue. We are very grateful to the following agencies and
organizations for financial support:
\end{sloppypar}

\begin{sloppypar}
Argentina -- Comisi\'on Nacional de Energ\'\i{}a At\'omica; Agencia Nacional de
Promoci\'on Cient\'\i{}fica y Tecnol\'ogica (ANPCyT); Consejo Nacional de
Investigaciones Cient\'\i{}ficas y T\'ecnicas (CONICET); Gobierno de la
Provincia de Mendoza; Municipalidad de Malarg\"ue; NDM Holdings and Valle
Las Le\~nas; in gratitude for their continuing cooperation over land
access; Australia -- the Australian Research Council; Belgium -- Fonds
de la Recherche Scientifique (FNRS); Research Foundation Flanders (FWO),
Marie Curie Action of the European Union Grant No.~101107047; Brazil --
Conselho Nacional de Desenvolvimento Cient\'\i{}fico e Tecnol\'ogico (CNPq);
Financiadora de Estudos e Projetos (FINEP); Funda\c{c}\~ao de Amparo \`a
Pesquisa do Estado de Rio de Janeiro (FAPERJ); S\~ao Paulo Research
Foundation (FAPESP) Grants No.~2019/10151-2, No.~2010/07359-6 and
No.~1999/05404-3; Minist\'erio da Ci\^encia, Tecnologia, Inova\c{c}\~oes e
Comunica\c{c}\~oes (MCTIC); Czech Republic -- GACR 24-13049S, CAS LQ100102401,
MEYS LM2023032, CZ.02.1.01/0.0/0.0/16{\textunderscore}013/0001402,
CZ.02.1.01/0.0/0.0/18{\textunderscore}046/0016010 and
CZ.02.1.01/0.0/0.0/17{\textunderscore}049/0008422 and CZ.02.01.01/00/22{\textunderscore}008/0004632;
France -- Centre de Calcul IN2P3/CNRS; Centre National de la Recherche
Scientifique (CNRS); Conseil R\'egional Ile-de-France; D\'epartement
Physique Nucl\'eaire et Corpusculaire (PNC-IN2P3/CNRS); D\'epartement
Sciences de l'Univers (SDU-INSU/CNRS); Institut Lagrange de Paris (ILP)
Grant No.~LABEX ANR-10-LABX-63 within the Investissements d'Avenir
Programme Grant No.~ANR-11-IDEX-0004-02; Germany -- Bundesministerium
f\"ur Bildung und Forschung (BMBF); Deutsche Forschungsgemeinschaft (DFG);
Finanzministerium Baden-W\"urttemberg; Helmholtz Alliance for
Astroparticle Physics (HAP); Helmholtz-Gemeinschaft Deutscher
Forschungszentren (HGF); Ministerium f\"ur Kultur und Wissenschaft des
Landes Nordrhein-Westfalen; Ministerium f\"ur Wissenschaft, Forschung und
Kunst des Landes Baden-W\"urttemberg; Italy -- Istituto Nazionale di
Fisica Nucleare (INFN); Istituto Nazionale di Astrofisica (INAF);
Ministero dell'Universit\`a e della Ricerca (MUR); CETEMPS Center of
Excellence; Ministero degli Affari Esteri (MAE), ICSC Centro Nazionale
di Ricerca in High Performance Computing, Big Data and Quantum
Computing, funded by European Union NextGenerationEU, reference code
CN{\textunderscore}00000013; M\'exico -- Consejo Nacional de Ciencia y Tecnolog\'\i{}a
(CONACYT) No.~167733; Universidad Nacional Aut\'onoma de M\'exico (UNAM);
PAPIIT DGAPA-UNAM; The Netherlands -- Ministry of Education, Culture and
Science; Netherlands Organisation for Scientific Research (NWO); Dutch
national e-infrastructure with the support of SURF Cooperative; Poland
-- Ministry of Education and Science, grants No.~DIR/WK/2018/11 and
2022/WK/12; National Science Centre, grants No.~2016/22/M/ST9/00198,
2016/23/B/ST9/01635, 2020/39/B/ST9/01398, and 2022/45/B/ST9/02163;
Portugal -- Portuguese national funds and FEDER funds within Programa
Operacional Factores de Competitividade through Funda\c{c}\~ao para a Ci\^encia
e a Tecnologia (COMPETE); Romania -- Ministry of Research, Innovation
and Digitization, CNCS-UEFISCDI, contract no.~30N/2023 under Romanian
National Core Program LAPLAS VII, grant no.~PN 23 21 01 02 and project
number PN-III-P1-1.1-TE-2021-0924/TE57/2022, within PNCDI III; Slovenia
-- Slovenian Research Agency, grants P1-0031, P1-0385, I0-0033, N1-0111;
Spain -- Ministerio de Ciencia e Innovaci\'on/Agencia Estatal de
Investigaci\'on (PID2019-105544GB-I00, PID2022-140510NB-I00 and
RYC2019-027017-I), Xunta de Galicia (CIGUS Network of Research Centers,
Consolidaci\'on 2021 GRC GI-2033, ED431C-2021/22 and ED431F-2022/15),
Junta de Andaluc\'\i{}a (SOMM17/6104/UGR and P18-FR-4314), and the European
Union (Marie Sklodowska-Curie 101065027 and ERDF); USA -- Department of
Energy, Contracts No.~DE-AC02-07CH11359, No.~DE-FR02-04ER41300,
No.~DE-FG02-99ER41107 and No.~DE-SC0011689; National Science Foundation,
Grant No.~0450696, and NSF-2013199; The Grainger Foundation; Marie
Curie-IRSES/EPLANET; European Particle Physics Latin American Network;
and UNESCO.
\end{sloppypar}

%
%
%

\end{document}